# Completely Enhanced Cell Phone Keypad

Md. Abul Kalam Azad, Rezwana Sharmeen, Shabbir Ahmad and S. M. Kamruzzaman [1]

Department of Computer Science & Engineering,

International Islamic University Chittagong,

Chittagong, Bangladesh.

[1] Department of Computer Science & Engineering,

Manarat International University,

Dhaka, Bangladesh.

Email: {azadarif, r_sharmin_79, bappi_51, smk_iiuc}@yahoo.com

**Abstract**

The enhanced frequency based keypad is designed to speed up the typing process. This paper will show that the proposed layout will increase the typing speed and be flexible for thumb. Traditional cell phone keypad is not a scientific keypad from the frequency point of view. Approaches have been explored to speed up the typing process. We found that no manufacturer has considered the frequency of the alphabet. The current architecture does not provide flexibility although the users are accustomed to the currently available multitapping keypad. Since the currently available keypad layouts are not best suited for users, this paper will suggest a keypad for cell phone and other cellular device based on the frequency of the alphabet in English language and also with the view of structure of human finger movements to provide a flexible and fast cell phone keypad. It also takes into consideration the key jamming problem that was available in typewriter. At first we identified those keys of cell phone, which are easily reachable and create less pressure on the thumb. Thus the key frequency order is calculated from anatomical point of view. In our proposed layout we arranged the alphabet in the frequent keys based on the frequency of the alphabet.

## 1. Introduction

WAP, chatting and messaging are considered to be the driving force of the next generation of mobile devices. New compact and extended function oriented cell phones from leading manufacturers are providing full pocket communication functionality. These cell phones allow users to browse the Internet, send and receive email, chatting and SMS, and handle personal data and information. All these services are closely related to text input or typing. However, one of the most important issues that could slow down or even prevent such devices from being widely used is the awkward user interface for text input. When SMS has become popular the 12-button alphanumeric keypad with traditional arrangement of alphanumeric characters was considered to be a major drawback. Many companies explored different approaches to solve this problem. We are proposing an approach to increase the typing speed thereby presenting a new mobile keypad layout taking into consideration the importance of mobile functionality in current world and making it more flexible to the users.

With the increase in the usability of cell phone, the manufacturers have designed different types of keypad layouts. Currently there are mainly two types of keypad layout one is Unitap and the other is Multitap. Unitap requires single tapping of a key for any character either numeric or alphabetic [4]. Manufacturers (Delta II, Blackberry) have designed micro sized QWERTY keypad, which is flexible for two hand operation [1,5]. Some company has developed Fastap architecture by arranging the numeric keys surrounded by the alphabetic keys. Multitap technology is the currently available most frequently used keypad layout. It requires multiple tapping for each character. It arranges the characters in alphabetic order. Each key has three or four characters. This type of layout is used by the leading manufacturers like Nokia, Samsung, Motorola, Erricson etc. Another type of keypad layout is designed known as MessagEase [13] based on Multitap technology but yet not implemented.

Taking into account the problems and the need of the cell phone user, we provide a new keypad layout with completely new arrangement of the alphabet which will make a radical change in the way text input to cell phone.

## 2. Proposed approach

We are proposing a new approach to speed up the typing and thereby eliminating the problem of key jamming by providing a new keypad layout, which will be flexible for user thumb. We have taken into consideration the concept of flexibility of thumb movement according to medical science and alphabetic frequency viewpoint with paying high consideration to jam protecting [2,3,5,6].

### 2.1. Categorizing the alphabet

We calculated the frequency of the appearance of alphabet in WAP, chatting and SMS purpose by using frequency-calculating algorithm. The most frequently used alphabet are shown in Table 1. In Table 1 Percentage refers to the percentage of occurrence of each alphabet in our sample inputs. In the below table "Alp" stands for Alphabet and "Per" stands for Percentage.

**Table 1:** Frequency of the single alphabetic character

| Alp | Per | Alp | Per | Alp | Per |
|---|---|---|---|---|---|
| e | 11.90% | l | 4.16% | p | 1.82% |
| t | 9.12% | d | 3.53% | b | 1.61% |
| o | 8.43% | u | 3.02% | v | 1.15% |
| a | 7.85% | m | 2.74% | k | 0.87% |
| i | 7.52% | y | 2.57% | j | 0.36% |
| s | 6.45% | c | 2.35% | x | 0.14% |
| n | 6.85% | w | 2.27% | q | 0.09% |
| r | 5.62% | f | 2.13% | z | 0.07% |
| h | 5.29% | g | 2.08% | | |

### 2.2. Identifying the frequently used keys

We proposed a new one-handed one fingered cell phone keypad layout. Our keypad layout is designed by arranging the most frequently used alphabet in the most frequently used keys. Frequent keys are those, which are flexible to the user. A key is flexible when it is reachable by the finger (thumb) without much pressurizing the physical structure and the internal joints of the thumb. A physical structure of the thumb is shown in Figure 1.

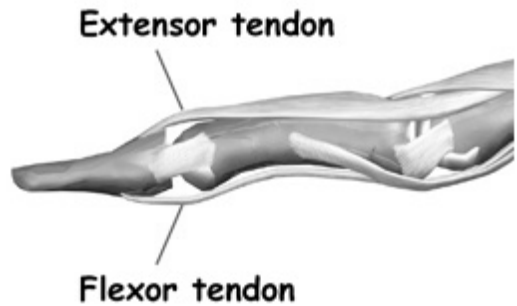

**Figure 1:** Physical structure of thumb joint

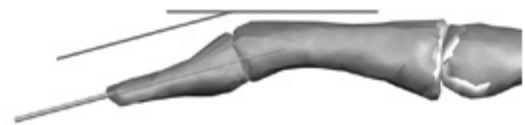

**Figure 2:** Physical structures of thumb joints in respect of angle

There are mainly two joints in a thumb (1) interphalangeal (2) metacarpophalangeal. Basically two types of thumb movements are required for pressing any key (1) Flexion (2) Extension. In case of Flexion movement the metacarpophalengeal joint is moved to forward direction whereas in case of extension metacarpophalengeal joint is moved to lateral direction, which is more painful and pressure creating. According to medical science, pressure on interphalangeal joint increases with the decrease in the joint angles. If the thumb movement pressurize the metacarpophalengeal joint to lateral direction then this movement pressurize the thumb and there by inconvenient and also a bit painful. The extension of the thumb is more painful and pressure creating than the decrease of angle in the interphalangeal joint [13,14,15]. Position of interphalangeal joint and metacarpophalangeal joint is shown in the following three figures: Figure 3, Figure 4 and Figure 5. In Figure 3 for pressing key 1 the interphalangeal joint is slightly bent and this angle of joint decreases as we proceed to press key 5 and key 9. While pressing key 9 there is extension of metacarpophalengeal to later direction.

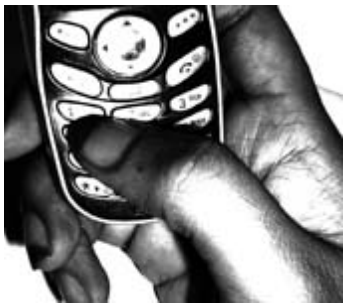

**Figure 3:** Pressing key 1

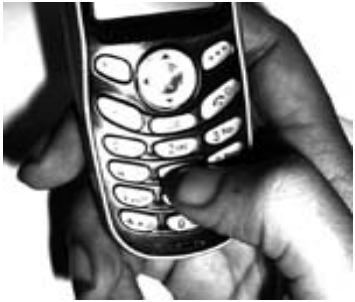

**Figure 4:** Pressing key 5

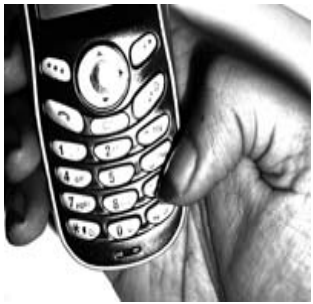

**Figure 5:** Pressing key 9

The pressure on the joints for thumb movements to press the cell phone keys are shown in Table 2. Where "IPJ" stands for Interphalangeal Joint, "MCJ" stands for Metacarpophalengeal Joint, "Flex" stands for Flexion and "Ext" stands for Extension.

**Table 2:** Statistical analysis of thumb movement for each key press

| Key | IPJ | MCJ | Flex | Ext |
|---|---|---|---|---|
| 1 | 120º | Forward direction | Yes | No |
| 2 | 110º | Forward direction | Yes | No |
| 3 | 80º | Lateral direction | No | Yes |
| 4 | 100º | Forward direction | Yes | No |
| 5 | 95º | Forward direction | Yes | No |
| 6 | 70º | Lateral direction | No | Yes |
| 7 | 80º | Forward direction | Yes | No |
| 8 | 70º | Lateral direction | No | Yes |
| 9 | 65º | Lateral direction | No | Yes |
| 0 | 40º | Lateral direction | No | Yes |

From the above table, the most frequently used key in the first row of keypad is 1 then 2 as the angle decreases to 10 degree. Also from the above table the most frequently used key in the second row is 4 then 5 as the angle decreases to 5 degree and the next frequent key in the third row is 7. The least frequent keys are consecutively 3, 6, 8, 9 and 0 as these movements require extension to lateral direction and also the angles in the interphalangeal joint is less. The final arrangement of frequency is 1 > 2 > 4 > 5 > 7 > 3 > 6 > 8 > 9 > 0.

### 2.3. Proposed keypad layout

Firstly we placed the most frequently used characters in the frequent keys in circular order from the most frequent key to least, then we again arranged the keys of next frequency from the least frequent key to most in reversed manner. Finally it was done again from most frequent key to least. In this way we designed our proposed keypad layout. The following table will show how the alphabet are arranged:

**Table 3:** Arrangement of single alphabet

| Key | 1 | 2 | 4 | 5 | 7 | 3 | 6 | 8 |
|---|---|---|---|---|---|---|---|---|
| Forward | e | t | o | a | i | s | n | r |
| Backward | w | c | y | m | u | d | l | h |
| Forward | f | g | p | b | n | k | jq | xz |

In our proposed new keypad layout the cell phone users will find the frequently used alphabet arranged in those keys, which are convenient to press and does not need to pay much pressure on the fingers. We have used key number 0 for blank space, which will also be convenient one. As the user will find out most of the frequent alphabet in easily reachable keys, this arrangement will speed up the typing process. In our proposed keypad layout the most inconvenient key is 9, which we have used for inserting symbols. The functions of the key 1 of traditional keypads will be shifted to key 9. It will lessen the finger works and will decrease the extra time and pressure required to move the position of the finger from the most frequently used key to the least frequent one. Hence these frequency based keypad will be very much flexible to the cell phone users in comparison to Unitap technology.

Our proposed keypad layout after arranging the alphabet is shown in the following, Figure 6.

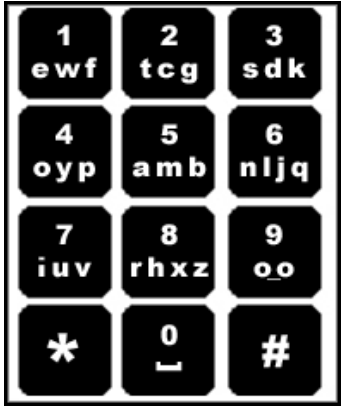

**Figure 6:** The proposed frequency based keypad layout

As we know key jamming is a severe problem in any sorts of keyboard or keypad and obviously in cell phone. To figure out the degree of key jamming we calculated the front combination of the alphabet of each key according two-alphabet combination and three-alphabet combination. We found that the top most frequent combination of two alphabet are shown in Table 3 where we found that the most frequent two-alphabet combination is 'am' which has the percentage of occurrence is only 1.11% and rest others are tend to zero. That's why it is apparent that this architecture is also key jamming proof. In the table below, "Syl" stands for Syllable and "Per" stands for Percentage.

**Table 4:** Frequency of the two character syllable

| Syl | Per | Syl | Per | Syl | Per |
|---|---|---|---|---|---|
| am | 1.11% | oy | 0.07% | lj | 0.00% |
| op | 0.78% | yp | 0.05% | tg | 0.00% |
| iv | 0.67% | nj | 0.03% | dk | 0.00% |
| ab | 0.57% | rh | 0.03% | lq | 0.00% |
| ef | 0.46% | iu | 0.02% | kq | 0.00% |
| nl | 0.33% | uv | 0.01% | rx | 0.00% |
| ew | 0.33% | nq | 0.01% | rz | 0.00% |
| mb | 0.31% | sd | 0.01% | hz | 0.00% |
| tc | 0.12% | cg | 0.01% | hz | 0.00% |
| sk | 0.11% | wf | 0.01% | xz | 0.00% |

## 3. Experimental Result

Key jamming problem has significant impact on typing speed. A perfect arrangement of key with the aim of speeding up the typing process, can dangerously hamper the typing speed if the key jamming problem is not taken into consideration. The early QWERTY keyboard is an eye opener; it had seriously taken into consideration the key-jamming problem, although two hands are simultaneously used. In case of cell phone we had to take this problem of key jamming into account seriously as only one hand and in most of the cases one finger are used. 'E' is the most frequently used alphabet and we placed it in key 1 the most frequently used key. Our next choice would have been 'T', the next frequently used alphabet. We placed it in key 2 rather than key 1 to protect key jamming. We arranged the frequently used alphabet in such a manner that the two consecutive alphabet will be low, such as 'ew' its frequency is 0.33% and the frequency of 'ef' is 0.46% and the frequency of 'wf' is 0.00%. Our arrangement is 'ewf', here the w character is placed to eliminate key jamming. Hence after pressing the alphabet of one key we have not to wait for the cursor to set, as the next character to be pressed will not be of that same key. Rather the next character will be in the next frequent key. By using this technology we tried to eliminate the most possible key jamming and therefore speeding up the typing process and not making the user to wait for the cursor to set.

We have an experiment on the data "the quick brown fox jumps over the lazy dog", and have calculated the number of times the most frequent keys and least frequent keys are used. This result is shown in Table 5:

**Table 5:** Quality comparison

| Technology | Use of most frequent keys | Use of least frequent keys |
|---|---|---|
| Traditional multitapping keypad | 10 | 24 |
| Proposed technology | 18 | 16 |

From the above table it is vivid that, in our proposed keypad layout is far more optimal then the traditional multitap keyboards. The user's thumb will be less pressurized as it will require the maximum use of frequent keys and the result will be significant increase in the typing process with less error rate.

We calculated the number of key pressing based on our test data. The result is shown in Table 6.

**Table 6:** Statistical comparison of key pressing

| Data | Current keypad | Proposed keypad |
|---|---|---|
| 148610 | 322780 | 218077 |
| 228632 | 496586 | 335504 |

The comparison of key pressing is shown in the following chart.

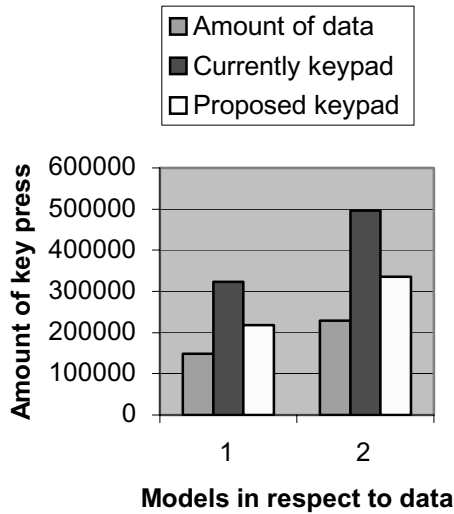

**Chart 1:** Statistical comparison of key pressing

From the above table it is vivid that the number of key pressing for the traditional Multitaping keypad is much higher than our proposed frequency based keypad layouts. As our keypad will require less key tapping hence it will be less time consuming and thereby increase the typing speed.

## 4. Conclusion

For speeding up the typing process the alphabetic arrangement are required to be changed to greater extent. In our proposed keypad layout as we have identified the frequently used keys and arranged the frequently used alphabet in those keys, it will be convenient and less pressure creating on the finger works. Although this process will require some learning time, the users will be benefited to great extent as soon as they become extended to it.

## 5. References

[1] For details of DELTA II technology: www.chicagologic.com
[2] For details of DVORAK technology: www.xpertkeyboard.com
[3] For details of PLUM keyboard: www.plum.bz
[4] For details of UNITAP technology: www.rl-technologies.com
[5] Historical background of QWERTY key board: www.ideafinder.com/history/inventions/story098.htm
[6] Historical background of DVORAK Keyboard www.dcn.davis.ca.us/~sander/mensa/dvorak1.html
[7] http://njnj.essortment.com/keyboardhisto_rdqo.htm
[8] www.webopedia.com
[9] www.wsdmag.com/Articles/ArticleID/5974/5974.html.
[10] Md. Hanif Seddiqui, Mohammad Mahadi Hassan, Md. Sazzad Hossain, Md. Nurul Islam "An optimal Keyboard Layout "in Proceedings of International Conference on Computer and Information Technology (ICCIT), Dhaka, Bangladesh, 2002.
[11] Mohammad Masud Hasan, Chowdhury Mofizur Rahman "Text Categorization Using Assocition Rule Based Decision Tree"in Proceedings of 6$^{th}$ International Conference on Computer and Information Technology (ICCIT), Dhaka, Bangladesh, 2003.
[12] Satish Narayana Srirama, Mohammad Abdullah Al Faruque and Mst Ayesha Siddika Munni"Alternatives to Mobile Keypad Design: Improved Text Feed "in Proceedings of 6$^{th}$ International Conference on Computer and Information Technology (ICCIT), Dhaka, Bangladesh, 2003.
[13] Scott MacKenzie, MessagEase multitap keypad: www.yorku.ca/mack/hcimobile02.html
[14] Structure of human thumb: www.anatomy-resources.com
[15] Structure of thumb joints: www.handuniversity.com/topics.asp?Topic_ID=19
[16] Structure of thumb: www.ncbi.nlm.nih.gov/entrez/